# PhaseStain: Digital staining of label-free quantitative phase microscopy images using deep learning

Yair Rivenson[1,2,3]†, Tairan Liu[1,2,3]†, Zhensong Wei[1,2,3]†, Yibo Zhang[1,2,3], Aydogan Ozcan[1,2,3,4]*

**Affiliations:**

[1]Electrical and Computer Engineering Department, University of California, Los Angeles, CA, 90095, USA.

[2]Bioengineering Department, University of California, Los Angeles, CA, 90095, USA.

[3]California NanoSystems Institute (CNSI), University of California, Los Angeles, CA, 90095, USA.

[4]Department of Surgery, David Geffen School of Medicine, University of California, Los Angeles, CA, 90095, USA.

*Correspondence: ozcan@ucla.edu

Address: 420 Westwood Plaza, Engineering IV Building, UCLA, Los Angeles, CA 90095, USA

Tel: +1(310)825-0915

Fax: +1(310)206-4685

†Equal contributing authors.

**ABSTRACT**

Using a deep neural network, we demonstrate a digital staining technique, which we term PhaseStain, to transform quantitative phase images (QPI) of label-free tissue sections into images that are equivalent to brightfield microscopy images of the same samples that are histochemically-stained. Through pairs of image data (QPI and the corresponding brightfield images, acquired after staining) we train a generative adversarial network (GAN) and demonstrate the effectiveness of this virtual staining approach using sections of human skin, kidney and liver tissue, matching the brightfield microscopy images of the same samples stained with H&E (Hematoxylin and Eosin), Jones' stain, and Masson's trichrome stain, respectively. This digital-staining framework might further strengthen various uses of label-free QPI techniques in pathology applications and biomedical research in general, by eliminating the need for chemical staining, reducing sample preparation related costs and saving time. Our results provide a powerful example of some of the unique opportunities created by data-driven image transformations enabled by deep learning.

## INTRODUCTION

Quantitative phase imaging (QPI) is a rapidly developing field, with a history of several decades in development[1,2]. QPI is a label-free imaging technique, which generates a quantitative image of the optical-path-delay through the specimen. Other than being label-free, QPI utilizes low-intensity illumination, while still allowing a rapid imaging time, which reduces phototoxicity in comparison to e.g., commonly-used fluorescence imaging modalities. QPI can be performed on multiple platforms and devices[3–7], from ultra-portable instruments all the way to custom-engineered systems integrated with standard microscopes, with different methods of extracting the quantitative phase information. QPI has also been recently used for the investigation of label-free thin tissue sections[2,8], which can be considered as a weakly scattering phase object, having limited amplitude contrast modulation under brightfield illumination.

Although QPI techniques result in quantitative contrast maps of label-free objects, the current clinical and research gold standard is still mostly based on brightfield imaging of histochemically labeled samples. The staining process dyes the specimen with colorimetric markers, revealing cellular and sub-cellular morphological information of the sample under brightfield microscopy. As an alternative, QPI has been demonstrated for the inference of local scattering coefficients of tissue samples[8,9]; for this information to be adopted as a diagnostic tool, some of the obstacles include the requirement of retraining experts and competing with a growing number of machine learning-based image analysis software[10,11], which utilize vast amounts of stained tissue images to perform e.g., automated diagnosis, image segmentation, or classification, among other tasks. One possible way to bridge the gap between QPI and standard image-based diagnostic modalities is to perform digital (i.e., virtual) staining of phase images of label-free samples to match the images of histochemically-stained samples. One previously used method for digital staining of tissue sections involves the acquisition of multi-modal, nonlinear microscopy images of the samples, while applying staining regents as part of the sample preparation, followed by a linear approximation of the absorption process to produce a pseudo-Hematoxylin and Eosin (H&E) image of the tissue section under investigation[12–14].

As an alternative to model-based approximations, deep learning has recently been successful in various computational tasks based on a data-driven approach, solving inverse problems in optics, such as super-resolution[15–17], holographic image reconstruction and phase recovery[18–21], tomography[22], Fourier ptychographic microscopy[23], localization microscopy[24–26] and ultra-short pulse reconstruction[27], among others. Recently, the application of deep learning for virtual staining of autofluorescence images of non-stained tissue samples has also been demonstrated[28]. Following on the success of these previous results, here we demonstrate that deep learning can be used for digital staining of label-free thin tissue sections using their quantitative phase images. For this image transformation between the phase image of a label-free sample and its stained brightfield image, which we term as PhaseStain, we used a deep neural network trained using the Generative Adversarial Network (GAN) framework[29]. Conceptually, PhaseStain (see Fig. 1) provides an image that is the digital equivalent of a brightfield image of the same sample after the chemical staining process; stated differently it transforms the phase image of a weakly scattering object (e.g., a label-free thin tissue section, which exhibits low amplitude modulation under visible light) into an amplitude object information, presenting the same color features that are observed under a brightfield microscope, after the chemical staining process.

We experimentally demonstrated the success of our PhaseStain approach using label-free sections of human skin, kidney and liver tissue that were imaged by a holographic microscope, matching the brightfield microscopy images of the same tissue sections stained with H&E, Jones' stain, and Masson's trichrome stain, respectively.

Deep learning-based virtual-staining of label-free tissue samples using quantitative phase images provide another important example of the unique opportunities enabled by data-driven image transformations. We believe that the PhaseStain framework will be instrumental for QPI community to further strengthen the uses of label-free QPI techniques for clinical applications and biomedical research, helping to eliminate the need for chemical staining, reduce sample preparation associated time, labor and related costs.

**RESULTS**

We trained 3 deep neural network models, which correspond to the 3 different combinations of tissue and stain types, i.e., H&E for skin tissue, Jones' stain for kidney tissue and Masson's trichrome for liver tissue. Following the training phase, these 3 trained deep networks were blindly tested on holographically reconstructed quantitative phase images (see the Methods section) that were not part of the network's training set. Figure 2 shows our results for virtual H&E staining of a phase image of a label-free skin tissue section, which confirms discohesive tumor cells lining papillary structures with dense fibrous cores. Additional results for virtual staining of quantitative phase images of label-free tissue sections are illustrated in Fig. 3, for kidney (digital Jones' staining) and liver (digital Masson's Trichrome staining). These virtually stained quantitative phase images show sheets of clear tumor cells arranged in small nests with a delicate capillary bed for the kidney tissue section, and a virtual trichrome stain highlighting normal liver architecture without significant fibrosis or inflammation, for the liver tissue section.

These deep learning-based virtual staining results presented in Figs. 2 and 3 visually demonstrate the high-fidelity performance of the GAN-based staining framework. To further shed light on this comparison between the PhaseStain results and the corresponding brightfield images of the chemically stained tissue samples, we quantified the structural similarity (SSIM) index of these two sets of images using:

$$SSIM(U_1, U_2) = \frac{1}{3} \sum_{i=1,2,3} \frac{(2\mu_{1,i}\mu_{2,i} + 2\sigma_{1,2,i} + c_2)}{\left(\mu_{1,i}^2 + \mu_{2,i}^2 + c_1\right)\left(\sigma_{1,i}^2 + \sigma_{2,i}^2 + c_2\right)} \quad (1)$$

where $U_1$, $U_2$ are the PhaseStain output and the corresponding brightfield reference image, respectively, $\mu_{k,i}$ and $\sigma_{k,i}$ are the mean and the standard deviation of each image $U_k$ ($k = 1,2$), respectively, and index $i$ refers to the RGB channels of the images. The cross-variance between the $i$-th image channels is denoted with $\sigma_{1,2,i}$ and $c_1$, $c_2$ are stabilization constants used to prevent division by a small denominator. The result of this analysis revealed that the SSIM was 0.8113, 0.8141 and 0.8905, for the virtual staining results corresponding to the skin, kidney and liver tissue samples, respectively, where the analysis was performed on ~10 Megapixel images, corresponding to a field-of-view (FOV) of ~1.47 mm$^2$ for each sample.

Next, to evaluate the sensitivity of the network output to phase noise in our measurements, we performed a numerical experiment on the quantitative phase image of a label-free skin tissue, where we added noise in the following manner:

$$\tilde{\phi}(m,n) = \phi(m,n) + \delta\phi(m,n) = \phi(m,n) + \beta r(m,n) * \frac{1}{2\pi L^2}\exp\left\{-(m^2+n^2)\Delta^2/\left[2(L\Delta)^2\right]\right\}, \qquad (2)$$

where $\tilde{\phi}$ is the resulting noisy phase distribution (i.e., the image under test), $\phi$ is the original phase image of the skin tissue sample, $r$ is drawn from a normal distribution $N(0,1)$, $\beta$ is the perturbation coefficient, $L$ is the Gaussian filter size/width and $\Delta$ is the pixel size, which spatially smoothens the random noise into isotropic patches, as shown in Fig. 4. We choose these parameters such that the overall phase signal-to-noise-ratio (SNR) is statistically identical for all the cases and made sure that no phase wrapping occurs. We then used 10 random realizations of this noisy phase image for 4 combinations of ($\beta$, $L$) values to generate $\tilde{\phi}$ which was used as input to our trained deep neural network.

The deep network inference fidelity for these noisy phase inputs is reported in Fig. 4, which reveals that it is indeed sensitive to local phase variations and related noise, and it improves its inference performance as we spatially extend the filter size, $L$ (while the SNR remains fixed). In other words, the PhaseStain network output is more impacted by small scale variations, corresponding to e.g., the information encoded in the morphology of the edges or refractive index discontinuities (or sharp gradients) of the sample. We also found that for a kernel size of $L\Delta$~3 μm, the SSIM remains unchanged (~0.8), across a wide range of perturbation coefficients, $\beta$. This result implies that the network is less sensitive to sample preparation imperfections, such as height variations and wrinkles in the thin tissue section, which naturally occur during the preparation of the tissue section.

## DISCUSSION

The training process of a PhaseStain network needs to be performed only once, following which, the newly acquired quantitative phase images of various samples are blindly fed to the pre-trained deep network to output a digitally-stained image for each label-free sample, corresponding to the image of the same sample FOV, as it would have been imaged with a brightfield microscope, following the chemical staining process. In terms of the computation speed, the virtual staining using PhaseStain takes 0.617 sec on average, using a standard desktop computer equipped with a dual-GPU for a FOV of ~0.45 mm$^2$, corresponding to ~3.22 Megapixels (see the implementation details in the Methods section). This fast inference time, even with relatively modest computers, means that the PhaseStain network can be easily integrated with a QPI-based whole slide scanner, since the network can output virtually-stained images in small patches while the tissue is still being scanned by an automated microscope, to simultaneously create label-free QPI and digitally-stained whole slide images of the samples.

The proposed technology has the potential to save time, labor and costs, by presenting an alternative to the standard histochemical staining workflow used in clinical pathology. As an example, one of the most common staining procedures (i.e., H&E stain) takes on average ~45 min and costs approximately $2-5, while the Masson's Trichrome staining procedure takes ~2-3 hours, with costs that range between $16-35, and often requires monitoring of the process by an expert, which is typically conducted by periodically examining the specimen under a microscope. In addition to saving time and costs, by circumventing the staining procedure, the tissue constituents would not be altered; this means the unlabeled tissue sections can be preserved for later analysis, such as matrix-assisted laser desorption ionization (MALDI) by micro-sectioning of specific areas[30] for molecular analysis or micro-marking of sub-regions which can be labeled with specific immunofluorescence tags or tested for personalized therapeutic strategies and drugs[31,32].

While in this study we trained 3 different neural network models to obtain optimal results for specific tissue and stain combinations, this does not pose a practical limitation for PhaseStain, since we can also train a more general digital staining model for a specific stain type (e.g. H&E, Jones' stain etc.) using multiple tissue types stained with it, at the cost of increasing the network size as well as the training and inference times[19]. Also, from clinical diagnostics perspective, the tissue type under investigation and the stain needed for its clinical examination are both known a priori, and therefore the selection of the correct neural network for each sample to be examined is straightforward to implement.

It is important to note that, in addition to the lensfree holographic microscope (see the Methods section) that we used in this work, PhaseStain framework can also be applied to virtually-stain the resulting images of various other QPI techniques, regardless of their imaging configuration, specific hardware or phase recovery method[2,6,7,33–36] that are employed.

One of the disadvantages of coherent imaging systems is "coherence-related image artifacts", such as e.g., speckle noise, or dust or other particles creating holographic interference fringes, which do not appear in incoherent brightfield microscopy images of the sample samples. In Fig. 5, we demonstrate the image distortions that, for example, out-of-focus particles create on the PhaseStain output image. To reduce such distortions in the network output images, coherence-related image artifacts resulting from out-of-focus particles can be digitally removed by using a recently introduced deep learning-based hologram reconstruction method, which learns, through data, to attack or eliminate twin-image artifact as well as interference fringes resulting from out-of-focus or undesired objects[19,20].

While in this manuscript we demonstrated the applicability of PhaseStain approach to fixed paraffin-embedded tissue specimen, our approach should be also applicable to frozen tissue sections, involving other tissue fixation methods as well (following a similar training process as detailed in the Methods section). Also, while our method was demonstrated for thin tissue sections, QPI has been shown to be valuable to image cells and smear samples (such as blood and Pap smears)[2,36], and PhaseStain technique would also be applicable to digitally stain these types of specimen.

To summarize, our presented results demonstrate some of the emerging opportunities created by deep learning for label-free quantitative phase imaging. The phase information resulting from various coherent imaging techniques can be used to generate a virtually stained image, translating the phase images of weakly scattering objects such as thin tissue sections into images that are equivalent to the brightfield images of the same samples, after the histochemical labeling. PhaseStain framework, in addition to saving time and cost associated with the labeling process, has the potential to further strengthen the use of label-free QPI techniques in clinical diagnostics workflow, while also preserving tissues for e.g., subsequent molecular and genetic analysis.

## MATERIALS AND METHODS

**Sample preparation and imaging**

All the samples that were used in this study were obtained from the Translational Pathology Core Laboratory (TPCL) and were prepared by the Histology Lab at UCLA. They were obtained after de-identification of the patient related information and were prepared from existing specimen. Therefore, this work did not interfere with standard practices of care or sample collection procedures.

Following formalin-fixing paraffin-embedding (FFPE), the tissue block is sectioned using a microtome into ~2-4 μm thick sections. This step is only needed for the training phase, where the transformation from a phase image into a brightfield image needs to be statistically learned. These tissue sections are then deparaffinized using Xylene and mounted on a standard glass slide using CytosealTM (Thermo-Fisher Scientific, Waltham, MA USA), followed by sealing of the specimen with a coverslip. In the learning/training process, this sealing step presents several advantages: protecting the sample during the imaging and sample handling processes, also reducing artifacts such as e.g., sample thickness variations.

Following the sample preparation, the specimen was imaged using an on-chip holographic microscope to generate a quantitative phase image (detailed in the next sub-section). Following the QPI process, the label-free specimen slide was put into Xylene for ~48 hours, until the coverslip can be removed without introducing distortions to the tissue. Once the coverslip is removed the slide was dipped multiple times in absolute alcohol, 95% alcohol and then washed in D.I. water for ~1 min. Following this step, the tissue slides were stained with H&E (skin tissue), Jones' stain (kidney tissue) and Masson's trichrome (liver tissue) and then coverslipped. These tissue samples were then imaged using a brightfield automated slide scanner microscope (Aperio AT, Leica Biosystems) with a 20×/0.75NA objective (Plan Apo), equipped with a 2× magnification adapter, which results an effective pixel size of ~0.25 μm.

**Quantitative phase imaging**

*Lensfree imaging setup:* Quantitative phase images of label-free tissue samples were acquired using an in-line lens-free holography setup[36]. A light source (WhiteLase Micro, NKT Photonics) with a center wavelength at 550 nm and a spectral bandwidth of ~2.5nm was used as the illumination source. The uncollimated light emitted from a single-mode fiber was used for creating a quasi-plane-wave that illuminated the sample. The sample was placed between the light source and the CMOS image sensor chip (IMX 081, Sony, pixel size of 1.12 μm) with a source-to-sample distance ($z_1$) of 5~10 cm and a sample-to-sensor distance ($z_2$) of 1-2 mm. This on-chip lensfree holographic microscope has submicron resolution with an effective pixel size of 0.37 μm, covering a sample FOV of ~20 mm$^2$ (which accounts for the entire active area of the sensor). The positioning stage (MAX606, Thorlabs, Inc.), that held the CMOS sensor, enabled 3D translation of the imager chip for performing pixel super-resolution (PSR)[5,36,37] and multi-height based iterative phase recovery[36,38]. All imaging hardware was controlled automatically by LabVIEW.

*Pixel super-resolution (PSR) technique:* To synthesize a high-resolution hologram (with a pixel size of ~0.37 μm) using only the G1 channel of the Bayer pattern (R, G1, G2, and B), a shift-and-add based PSR algorithm was applied[39,37]. The translation stage that holds the image sensor was programmed to laterally shift on a 6×6 grid with sub-pixel spacing at each sample-to-sensor distance. A low-resolution hologram was recorded at each position and the lateral shifts were precisely estimated using a shift estimation algorithm[36]. This step results in 6 non-overlapping panels that were each padded to a size of 4096×4096 pixels, and were individually phase-recovered, which is detailed next.

*Multi-height phase recovery:* Lensfree in-line holograms at eight sample-to-sensor distances were captured. The axial scanning step size was chosen to be 15 μm. Accurate z-steps were obtained by applying a holographic autofocusing algorithm based on the edge sparsity criterion ("Tamura of the gradient", i.e., ToG)[40]. A zero-phase was assigned to the object intensity measurement as an initial phase guess, to start the iterations. An iterative multi-height phase recovery algorithm[41] was then used by propagating the complex field back and forth between each height using the transfer function of free-space[42]. During this iterative process, the phase was kept unchanged at each axial plane, where the amplitude was updated by using the square-root of the object intensity measurement. One iteration was defined as propagating the hologram from the 8$^{th}$ height (farthest from the sensor chip) to the 1st height (nearest to the sensor) then back propagating the complex field to the 8$^{th}$ height. Typically, after 10-30 iterations the phase is retrieved. For the final step of the reconstruction, the complex wave defined by the converged amplitude and phase at a given hologram plane was propagated to the object plane[42], from which the phase component of the sample was extracted.

**Data preprocessing and image registration**

An important step in our training process is to perform an accurate image registration, between the two imaging modalities (QPI and brightfield), which involves both global matching and local alignment steps. Since the network aims to learn the transformation from a label-free phase retrieved image to a histochemically-stained brightfield image, it is crucial to accurately align the FOVs for each input and target image pair in the dataset. We perform this cross-modality alignment procedure in four steps; steps 1,2 and 4 are done in Matlab (The MathWorks Inc., Natick, MA, USA) and step 3 involves TensorFlow.

The first step is to find a roughly matched FOV between QPI and the corresponding brightfield image. This is done by first bicubic down-sampling the whole slide image (WSI) (~60k by 60k pixels) to match the pixel size of the phase retrieved image. Then, each 4096×4096-pixel phase image was cropped by 256 on each side (resulting in an image with 3584×3584 pixels) to remove the padding that is used for the image reconstruction process. Following this step, both the brightfield and the corresponding phase images are edge extracted using the Canny method[43], which uses double threshold to detect strong and weak edges on the gradient of the image. Then, a correlation score matrix is calculated by correlating each 3584x3584-pixel patch of the resulting edge image to the same size as the image extracted from the brightfield edge image. The image with the highest correlation score indicates a match between the two images, and the corresponding brightfield image is cropped out from the WSI. Following this initial matching procedure, the quantitative phase image and the brightfield microscope images are coarsely matched.

The second step is used to correct for potential rotations between these coarsely matched image pairs, which might be caused by a slight mismatch in the sample placement during the two image acquisition experiments (which are performed on different imaging systems, holographic vs. brightfield). This intensity-based registration step correlates the spatial patterns between the two images; phase image that is converted to unsigned integer format and the luminance component of the brightfield image were used for this multimodal registration framework implemented in Matlab. The result of this digital procedure is an affine transformation matrix, which is applied to the brightfield microscope image patch, to match it with the quantitative phase image of the same sample. Following this registration step, the phase and the corresponding brightfield images are globally aligned. A further crop of 64 pixels on each side to the aligned image pairs is used to accommodate for a possible rotation angle correction.

The third step involves the training of a separate neural network that roughly learns the transformation from quantitative phase images into stained brightfield images, which can help the distortion correction between the two image modalities in the fourth/final step. This neural network has the same structure as the network that was used for the final training process (see the next sub-section on GAN architecture and its training) with the input and target images obtained from the second registration step discussed earlier. Since the image pairs are not well aligned yet, the training is stopped early at only ~2000 iterations to avoid a structural change at the output to be learnt by the network. The output and target images of the network are then used as the registration pairs in the fourth step, which is an elastic image registration algorithm, used to correct for local feature registration[16].

**GAN architecture and training**

The GAN architecture that we used for PhaseStain is detailed in Table 1. Following the registration of the label-free quantitative phase images to the brightfield images of the histochemically stained tissue sections, these accurately aligned fields-of-view were partitioned to overlapping patches of 256×256 pixels, which were then used to train the GAN model. The GAN is composed of two deep neural networks, a generator and a discriminator. The discriminator network's loss function is given by:

$$\ell_{discrimnator} = D(G(x_{input}))^2 + (1 - D(z_{label}))^2 \qquad (3)$$

where $D(.)$ and $G(.)$ refer to the discriminator and generator network operator, $x_{input}$ denotes the input to the generator, which is the label-free quantitative phase image, and $z_{label}$ denotes the brightfield image of the chemically stained tissue. The generator network, G, tries to generate an output image with the same statistical features as $z_{label}$, while the discriminator, D, attempts to distinguish between the target and the generator output images. The ideal outcome (or state of equilibrium) will be when the generator's output and target images share an identical statistical distribution, where in this case, $D(G(x_{input}))$ should converge to 0.5. For the generator deep network, we defined the loss function as:

$$\ell_{generator} = L_1\{z_{label}, G(x_{input})\} + \lambda \times TV\{G(x_{input})\} + \alpha \times (1 - D(G(x_{input})))^2 \qquad (4)$$

where $L_1\{.\}$ term refers to the absolute pixel by pixel difference between the generator output image and its target, $TV\{.\}$ stands for the total variation regularization that is being applied to the generator output, and the last term reflects a penalty related to the discriminator network prediction of the generator output.

The regularization parameters (λ, α) were set to 0.02 and 2000 so that the total variation loss term, $\lambda \times \text{TV}\{G(x_{input})\}$, was ~2% of the $L_1$ loss term, and the discriminator loss term, $\alpha \times (1 - D(G(x_{input})))^2$ was ~98% of the total generator loss, $\ell_{generator}$.

For the generator deep neural network, we adapted the U-net architecture[44], which consists of a down-sampling and an up-sampling path, with each path containing 4 blocks forming 4 distinct levels (see Table 1). In the down-sampling path, each residual block consists of 3 convolutional layers and 3 leaky rectified linear (LReLU) units used as an activation function, which is defined as:

$$\text{LReLU}(x) = \begin{cases} x & \text{for } x > 0 \\ 0.1x & \text{otherwise} \end{cases} \qquad (5)$$

At the output of each block, the number of channels is 2-fold increased (except for the first block that increases from 1 input channel to 64 channels). Blocks are connected by an average-pooling layer of stride 2 that down-samples the output of the previous block by a factor of 2 for both horizontal and vertical dimensions (as shown in Table 1).

In the up-sampling path, each block also consists of 3 convolutional layers and 3 LReLU activation functions, which decrease the number of channels at its output by 4-fold. Blocks are connected by a bilinear up-sampling layer that up-samples the size of the output from the previous block by a factor of 2 for both lateral dimensions. A concatenation function with the corresponding feature map from the down-sampling path of the same level is used to increase the number of channels from the output of the previous block by 2. The two paths are connected in the first level of the network by a convolutional layer which maintains the number of the feature maps from the output of the last residual block in the down-sampling path (see Table 1). The last layer is a convolutional layer that maps the output of the up-sampling path into 3 channels of the YCbCr color map.

The discriminator network consists of one convolutional layer, 5 discriminator blocks, an average pooling layer and two fully connected layers. The first convolutional layer receives 3 channels (YCbCr color map) from either the generator output or the target, and increases the number of channels to 64. The discriminator blocks consist of 2 convolutional layers with the first layer maintaining the size of the feature map and the number of channels, while the second layer increases the number of channels by 2-fold and decreases the size of the feature map by 4-fold. The average pooling layer has a filter size of 8×8, which results in a matrix with a size of ($B$, 2048), where $B$ refers to the batch size. The output of this average pooling layer is then fed into two fully connected layers with the first layer maintaining the size of the feature map, while the second layer decreases the output channel to 1, resulting in an output size of ($B$, 1). The output of this fully connected layer is going through a sigmoid function indicating the probability that the 3-channel discriminator input is drawn from a chemically stained brightfield image. For the discriminator network, all the convolutional layers and fully connected layers are connected by LReLU nonlinear activation functions.

Throughout the training, the convolution filter size was set to be 3×3. For the patch generation, we applied data augmentation by using 50% patch overlap for the liver and skin tissue images, and 25% patch overlap for the kidney tissue images (see Table 2). The learnable parameters including filters, weights and biases in the convolutional layers and fully connected layers are updated using an adaptive moment estimation (Adam) optimizer with learning rate $1\times10^{-4}$ for the generator network and $1\times10^{-5}$ for the discriminator network.

For each iteration of the discriminator, there were $v$ iterations of the generator network; for liver and skin tissue training, $v = \max(5, floor(7-w/2))$ where we increased $w$ by 1 for every 500 iterations ($w$ was initialized as 0). For the kidney tissue training, we used $v = \max(4, floor(6-w/2))$ where we increased $w$ by 1 for every 400 iteration. This helped us to train the discriminator not to overfit to the target brightfield images. We used a batch size of 10 for the training of liver and skin tissue sections, and 5 for the kidney tissue sections. The network's training stopped when the validation set's $L_1$-loss did not decrease after 4000 iterations. A typical convergence plot of our training is shown in Fig. 6.

**Implementation details**

The number of image patches that were used for training, the number of epochs and the training schedules are shown in Table 2. The network was implemented using Python version 3.5.0, with TensorFlow framework version 1.7.0. We implemented the software on a desktop computer with a Core i7-7700K CPU @ 4.2GHz (Intel) and 64GB of RAM, running a Windows 10 operating system (Microsoft). Following the training for each tissue section, the corresponding network was tested with 4 image patches of 1792×1792 pixels with an overlap of ~7%. The outputs of the network were then stitched to form the final network output image of 3456×3456 pixels (FOV ~1.7 mm$^2$), as shown in e.g., Fig. 2. The network training and testing were performed using dual GeForce GTX 1080Ti GPUs (NVidia).

**ACKNOWLEDGMENT**

The Ozcan Research Group at UCLA acknowledges the support of NSF Engineering Research Center (ERC, PATHS-UP), the Army Research Office (ARO; W911NF-13-1-0419 and W911NF-13-1-0197), the ARO Life Sciences Division, the National Science Foundation (NSF) CBET Division Biophotonics Program, the NSF Emerging Frontiers in Research and Innovation (EFRI) Award, the NSF INSPIRE Award, NSF Partnerships for Innovation: Building Innovation Capacity (PFI:BIC) Program, the National Institutes of Health (NIH, R21EB023115), the Howard Hughes Medical Institute (HHMI), Vodafone Americas Foundation, the Mary Kay Foundation, and Steven & Alexandra Cohen Foundation. Yair Rivenson is partially supported by the European Union's Horizon 2020 research and innovation programme under the Marie Skłodowska-Curie grant agreement No H2020-MSCA-IF-2014-659595 (MCMQCT). The authors also acknowledge the Translational Pathology Core Laboratory (TPCL) and the Histology Lab at UCLA for their assistance with the sample preparation and staining, as well as Prof. Dean Wallace of Pathology and Laboratory Medicine at UCLA's David Geffen School of Medicine for image evaluations.

**FIGURES AND TABLES**

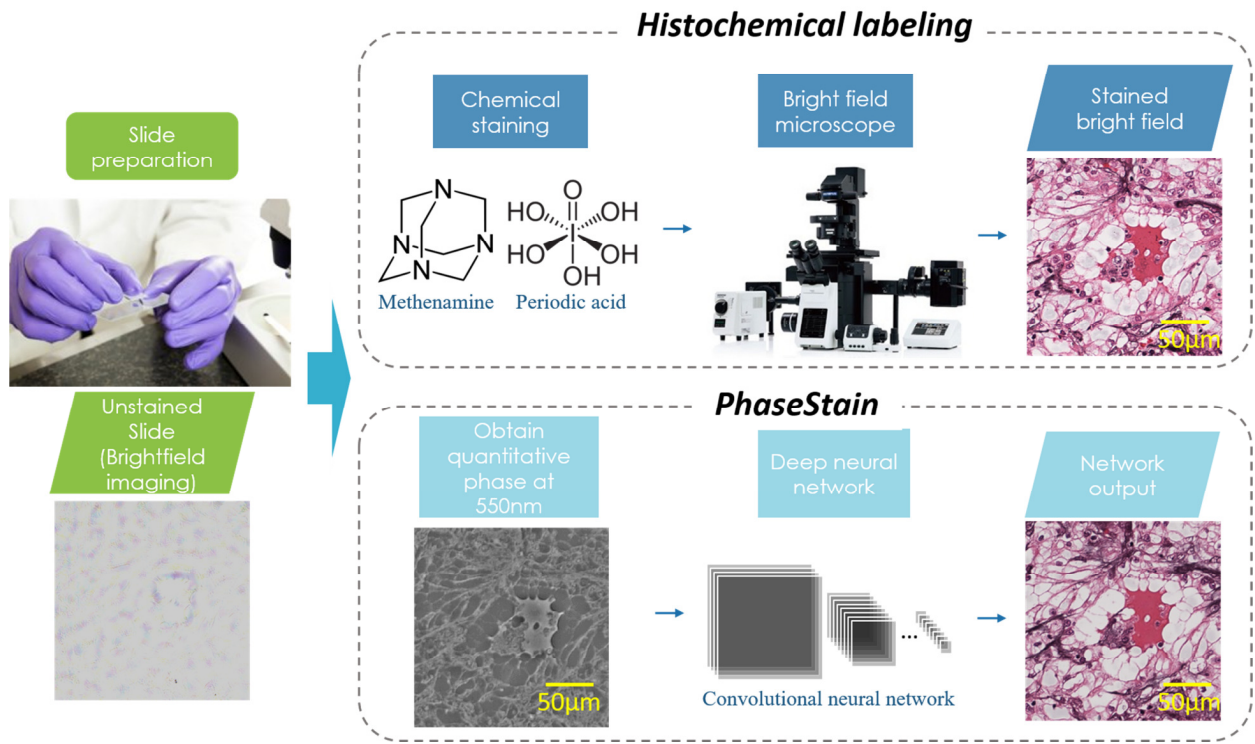

**Fig. 1.** PhaseStain workflow: Quantitative phase image of a label-free specimen is virtually stained by a deep neural network, bypassing the standard histochemical staining procedure that is used as part of clinical pathology.

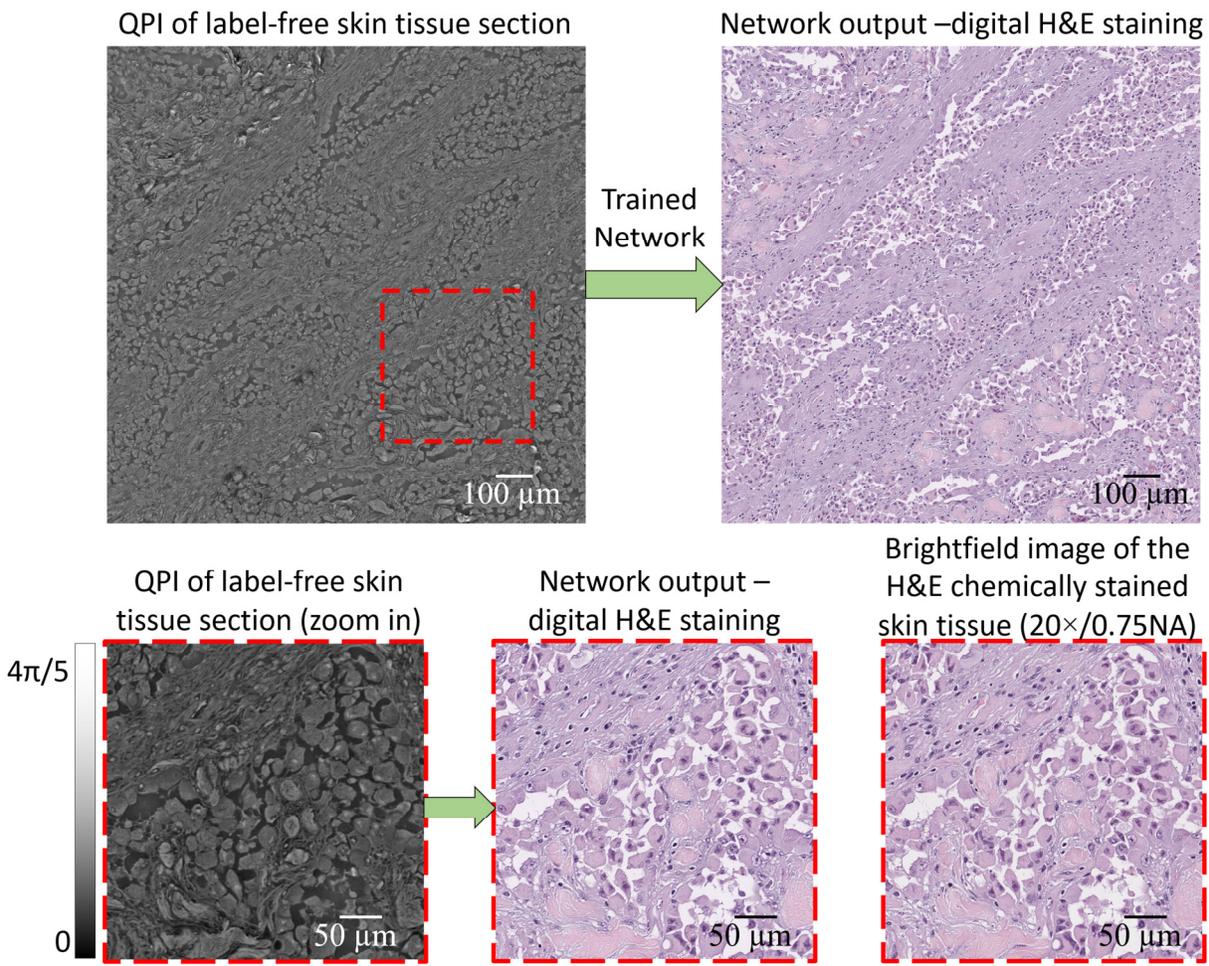

**Fig. 2.** Virtual H&E staining of label-free skin tissue using PhaseStain framework.

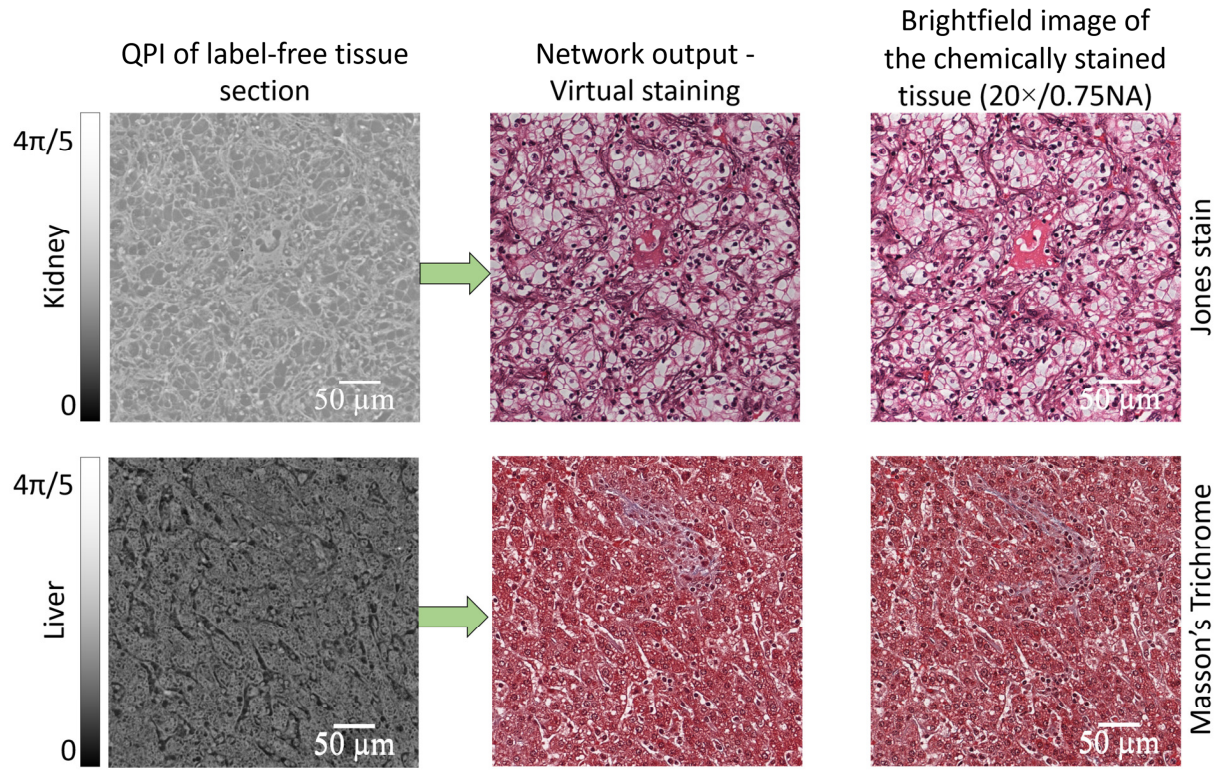

**Fig. 3.** PhaseStain based virtual staining of label-free kidney tissue (Jones' stain) and liver tissue (Masson's Trichrome).

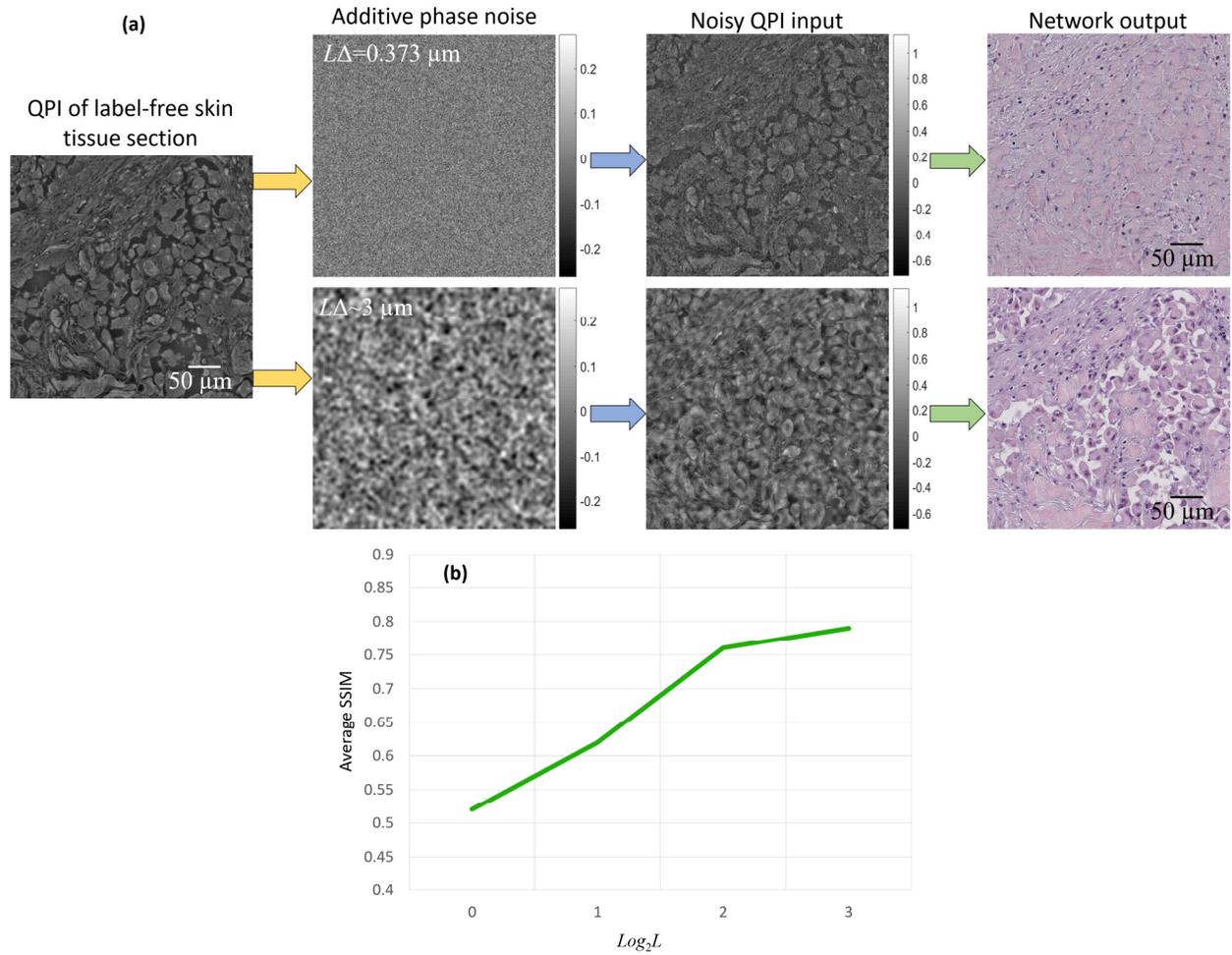

**Fig. 4.** (a) PhaseStain results for noisy phase input images (ground truth shown in Fig. 2). Top row: $L\Delta \sim 0.373$ µm; second row: $L\Delta \sim 3$ µm. (b) Analysis of the impact of phase noise on the inference quality of PhaseStain (quantified using SSIM), as a function of the Gaussian filter length, $L$ (see Eq. 2).

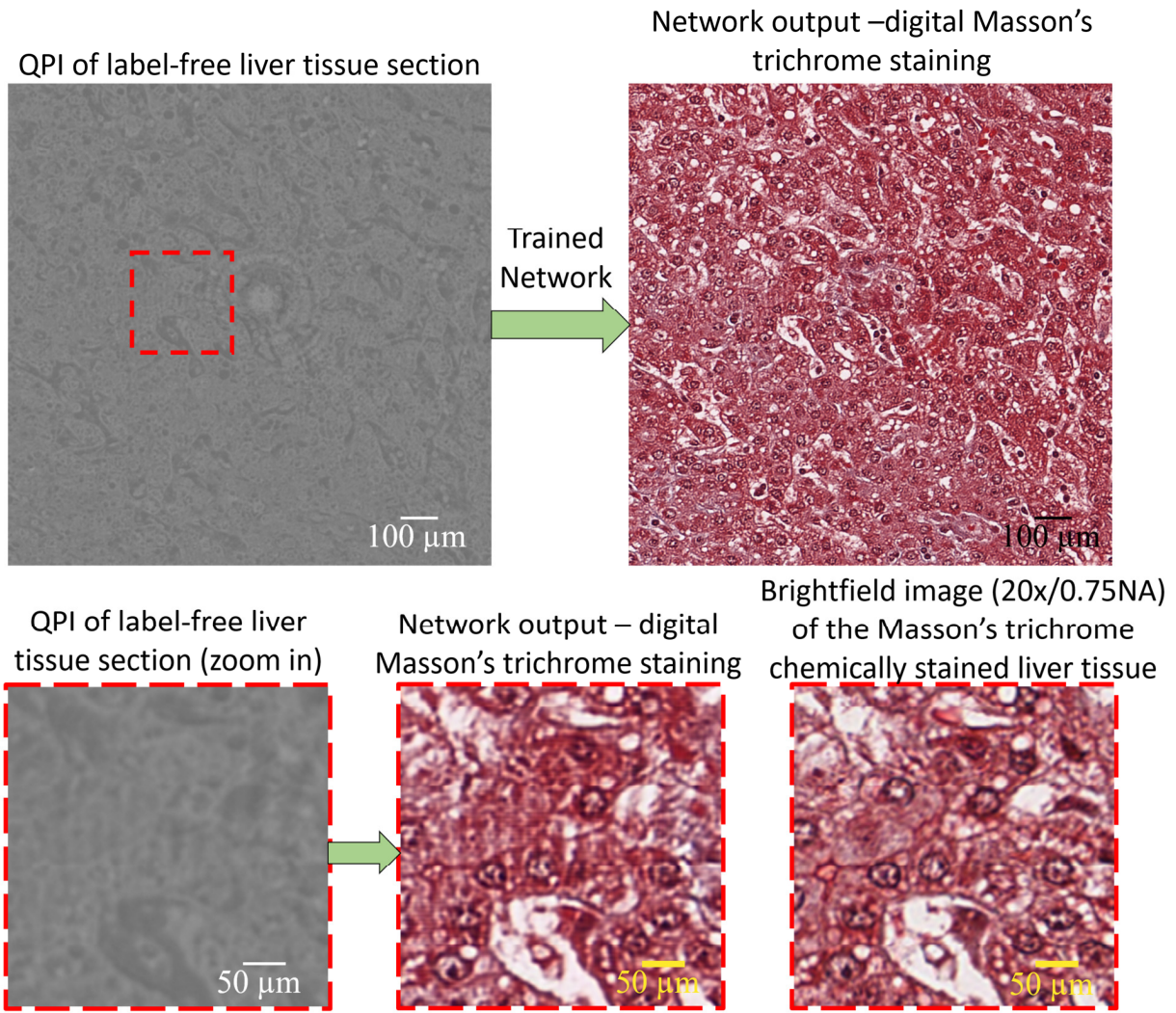

**Fig. 5.** The impact of holographic fringes resulting from out-of-focus particles on the deep neural network's digital staining performance.

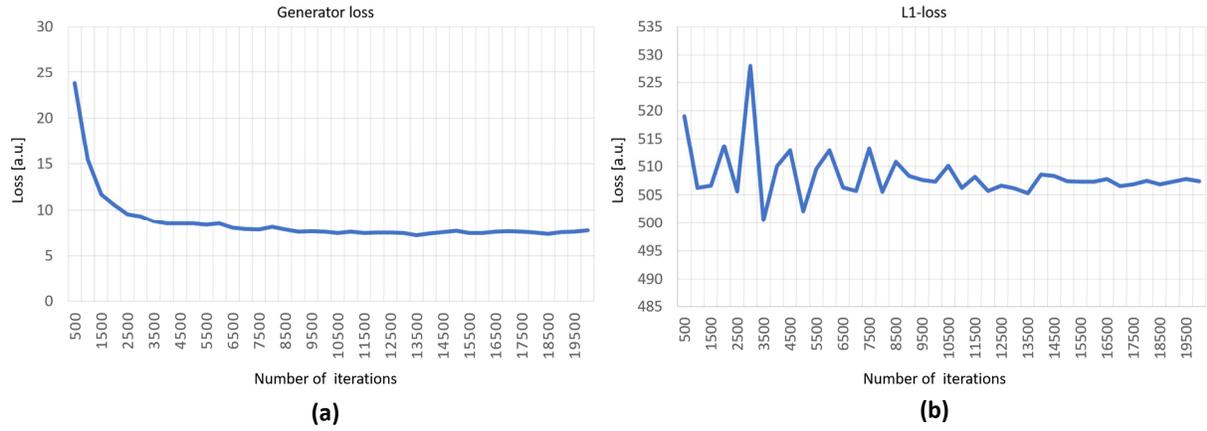

**Fig. 6.** PhaseStain convergence plots for the validation set of the digital H&E staining of the skin tissue. (a) $L1$-loss with respect to the number of iterations. (b) Generator loss, $\ell_{\text{generator}}$ with respect to the number of iterations.

**(a) *Generator***

Down path

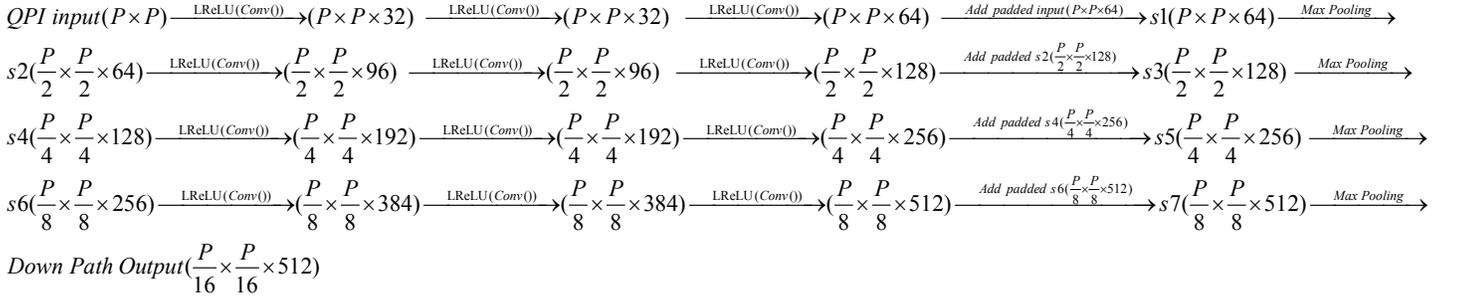

Connection layer

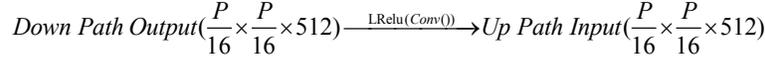

Up path

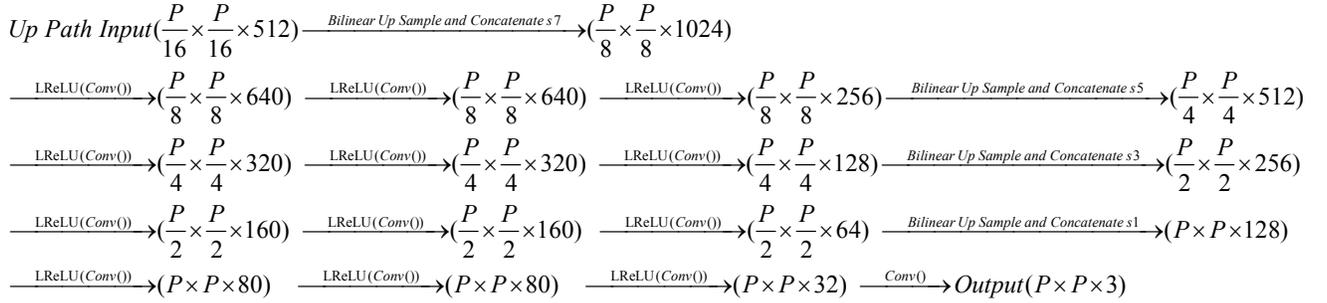

**(b) *Discriminator***

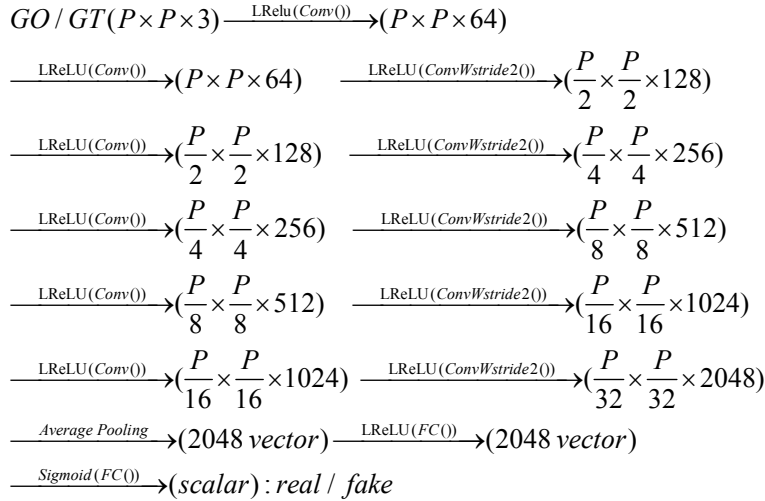

**Table 1.** The GAN architecture. LReLU: Leaky ReLU, Conv: convolutional layer with a stride of 1, FC: fully-connected layer, ConvWstride2: convolutional layer with stride 2, GO: generated output, GT: ground truth image.

| Tissue type | # of iterations | # of patches (256×256 pixels) | Training time (hours) | # of epochs |
|---|---|---|---|---|
| Liver | 7500 | 2500 training / 625 validation | 11.076 | 25 |
| Skin | 11000 | 2500 training / 625 validation | 11.188 | 18 |
| Kidney | 13600 | 2312 training / 578 validation | 13.173 | 39 |

**Table 2.** Training details for virtual staining of different tissue types using PhaseStain. Following the training, blind inference takes ~0.617 s for a FOV of ~0.45 mm$^2$, corresponding to ~3.22 Megapixels (see the Discussion section).